# Enabling Shortwave-QKD in Short-Reach Networks: Impact of a Composite ODN Native to Telecom Applications

*Mariana F. Ramos,[†] Costin Luchian, Michael Hentschel, Florian Honz, Marie-Christine Slater, Hannes Hübel, and Bernhard Schrenk*
AIT Austrian Institute of Technology, Center for Digital Safety & Security / Security & Communication Technologies, 1210 Vienna, Austria

[†]marianarms.13@gmail.com

**Abstract:** We deploy shortwave-QKD over short-reach in-house/datacom architectures and show that few-mode propagation and speckle-selective loss severely impact the QKD performance. We accomplish 12 kb/s secure-key generation in presence of 50 co-existing data channels.

## 1. Introduction

The increasing demand for data-intensive applications and cloud computing puts significant strain to intra-datacenter communication. As datacenters expand, secure high-capacity interconnects become critical. Quantum key distribution (QKD) has emerged as a promising solution to enhance data security [1]. Although longer wavelengths are commonly employed in QKD systems, the 850-nm band offers distinct advantages for short-range applications: : its large spectral separation from telecom bands reduces in-band Raman and nonlinear crosstalk from coexisting classical channels, and it enables the use of mature silicon single-photon avalanche diodes (Si-SPADs) with high detection efficiency, low dark counts at or near room temperature—facilitating compact, CMOS-compatible receiver integration and lowering system cost. This includes compatibility with industrial facilities or datacenter architectures where ~90% of communication links are shorter than 350 m [2].

In this work, we advance our earlier work [3] by accounting for speckle-dependent impairments over planar light wave circuit (PLC) splitter technology. PLC refers here to a glass/silica-based integrated-optics device implementing passive functions (e.g., splitters), a well-established component technology in classical telecommunications. We prove the feasibility of deploying shortwave-QKD within branched network architectures by analyzing various optical distribution networks (ODN), including straight-line and tree-shaped configurations (Fig. 1). The ODN refers to the passive fiber infrastructure (splitters, and distribution/drop fibers) between the transmitter and the receiver, as illustrated in Fig. 1. We show suitable mitigation methods for few-mode propagation of a shortwave-QKD channel (employing the BB84 protocol) over standard telecom components, as they result from the optical mode mismatch [4,5]. Towards that, we demonstrate secure-key generation in co-existence with 50 classical channels and operation over field-installed fiber. Complementing our short-reach focus, recent field trials have demonstrated robust QKD–classical coexistence at very high capacities, including successful QKD operation over 26 km of field-deployed multicore fiber carrying 110 Tb/s of classical traffic [9], thereby situating our work within the broader coexistence landscape.

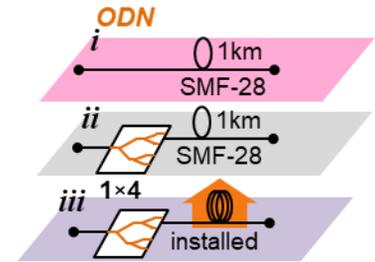

Fig. 1. ODN layouts with few-mode propagation.

Table 1. Evaluated ODN configurations.

| Scenario | | 1×4 split | M1 | 1km drop | M2 |
|---|---|---|---|---|---|
| 1✶ | straight line | | | ✓ | ✓ |
| 2■ | colorless distribution (tree) | ✓ | ✓ | ✓ | |
| 3▲ | | ✓ | | ✓ | ✓ |
| 4● | | ✓ | ✓ | ✓ | ✓ |
| 5▲ | co-existence | ✓ | | ✓ | ✓ |
| 6▲ | field-installed | ✓ | | ✓ | ✓ |

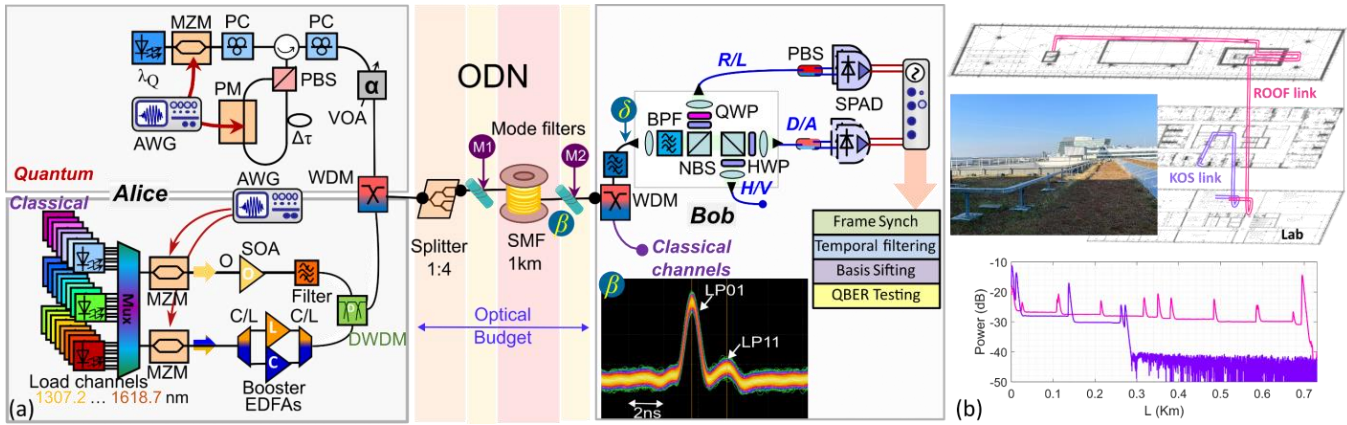

Fig. 2. (a) Setup for shortwave QKD. (b) Field-installed in-door office (KOS) and roof-top (ROOF) fiber segments and OTDR traces.

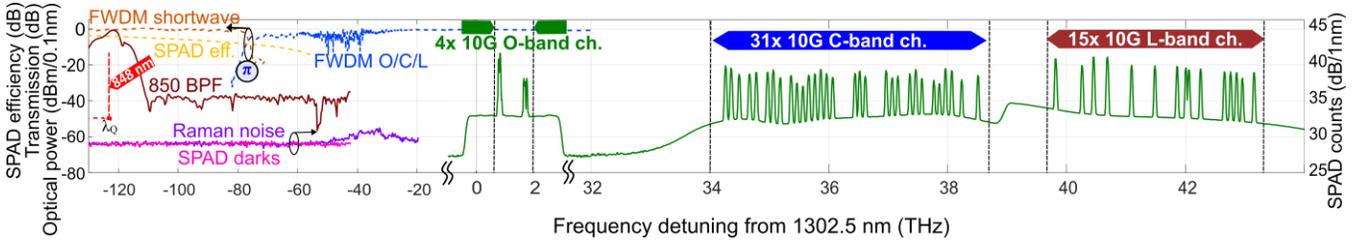

Fig. 3. Shortwave spectra with co-existing classical channels in the O-, C- and L- bands, filter transmission and SPAD efficiency.

## 2. Shortwave-QKD in Telecom Fiber Plants

The proposed QKD integration scheme capitalizes on the wide spectral detuning to telecom windows in the O-, C- and L-bands when populating the shortwave band with a QKD channel. The network scenarios summarized in Table 1 consider signal distribution through 1×N splitters and branching fibers that are designed for single-mode operation in the 2nd/3rd telecom windows. With this, the potential benefits of shortwave-QKD are opposed by two challenges: (i) facing a higher fiber loss of 1.8 dB/km and (ii) few-mode propagation over ITU-T G.652B-type SMF with a cut-off wavelength of 1260 nm [3,6], which together with the polarization-encoded QKD signal can lead to depolarization. Especially the second aspect requires careful attention: when operating below the cut-off, the amount of light transmitted through any optical discontinuity depends on the exact positioning of the few-mode profile and the portion that aligns with the overlap of fiber cores [7]. Moreover, the distribution of optical power among drop ports of commonly used fiber-optic distribution elements, such as PLC splitters, can exhibit fluctuations as a response to changes in the wavelength-selective speckle [4]. We will therefore employ a combination of spectral launch control and mode filtering to address few-mode QKD operation over standard telecom fiber (SMF28).

Figure 2 presents the experimental setup for investigating shortwave-QKD over various ODN configurations. Alice encodes four polarization states (A/D and R/L) at a wavelength of $\lambda_Q$ = 848 nm according to the method presented in [3,8]. The symbol rate for two sequentially encoded polarization states is 445 MHz. The launch power is set for an average photon number of $\mu$ = 0.1 hv/symbol. Bob decodes the polarization states and measures them in both bases, employing silicon SPADs with a detection efficiency of 38% and an intrinsic dark count rate of 350 cts/s. Detection events are time-tagged and passed to a real-time processing unit, which handles frame synchronization, temporal filtering at 20% of the symbol width, and basis sifting prior to QBER estimation. The short-reach ODN between Alice and Bob involves a 1-km long SMF28, a 1×4 PLC-based power splitter with SMF28 pigtails, and two possible locations for the mode filter (M1 and M2) to mitigate few-mode propagation artifacts. The launched mode of the shortwave-QKD signal evolves in both LP01 and LP11 modes after few-mode fiber propagation, as shown in the inset β in Fig. 1a. To remove the LP11 mode that is subject to a differential mode delay of 2.02 ns/km, which would lead to reception penalties due to modal dispersion and, specifically, depolarization, a simple mode filter made of 3 fiber loops (mandrel diameter of 2.1 cm) of SMF28 can be included after each joint.

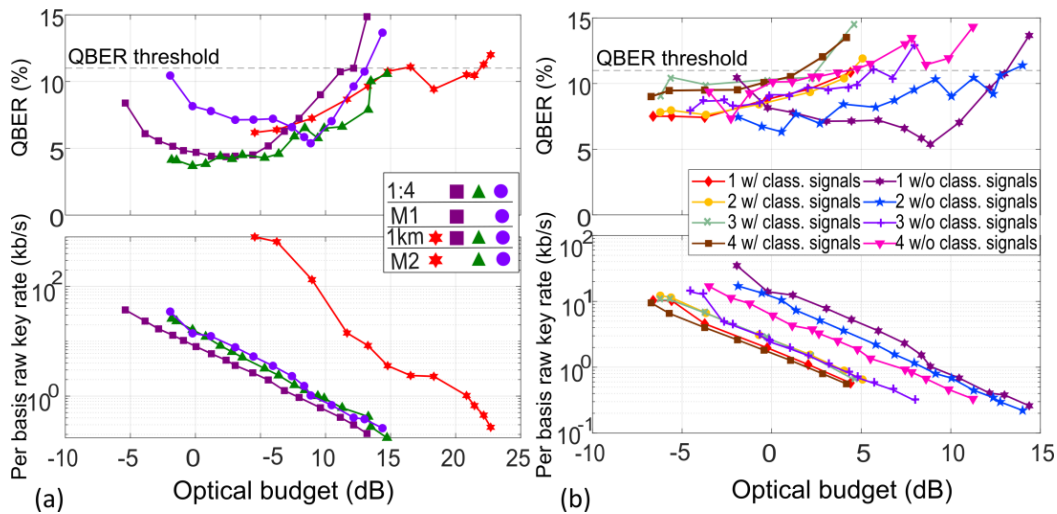

Fig. 4. Raw-key rate and QBER as a function of OB (a) for different scenarios configuration, and (b) in co-existence with classical signals for different splitter outputs.

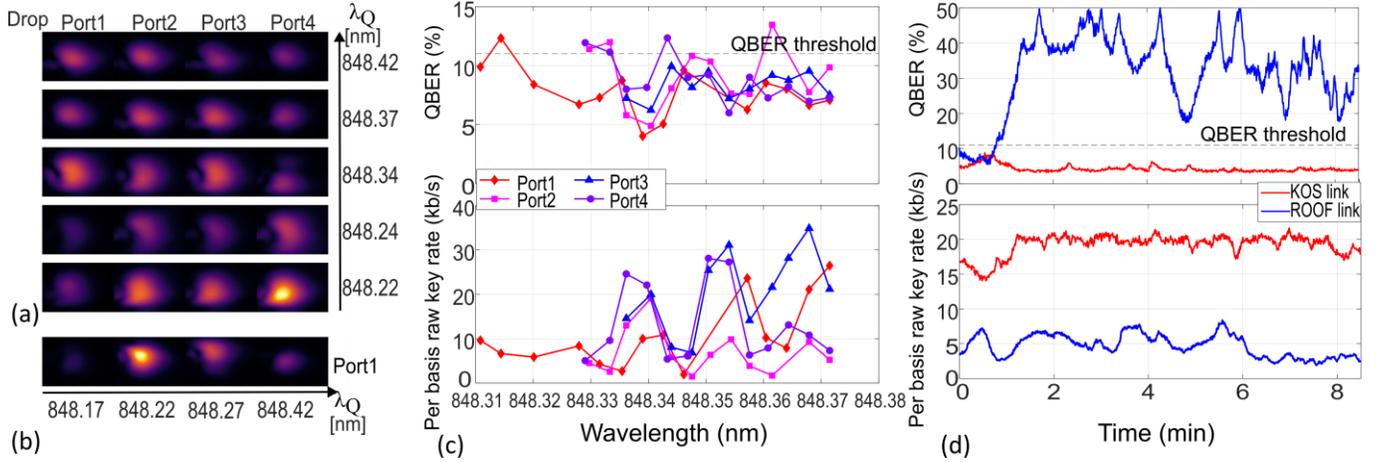

Fig. 5. Speckle after transmission over a 1×4 PLC splitter (a) for all 4 ports when detuning $\lambda_Q$, and (b) specifically for port 1. (c) Key rate and QBER as function of wavelength detuning. (d) Time evolution of QBER and key rate for field-installed links.

### 3. Shortwave Few-Mode QKD Propagation

Figure 4a shows the raw-key rate and QBER as function of optical link budget (OB) for different ODN configurations. The first (✱) includes a straight 1-km link of SMF28 and M2. It allows for the highest key rate (800 kb/s per basis) and a maximum channel loss of 22 dB. The migration to a signal-distributing ODN based on an optical splitter significantly worsens the conditions for QKD operation: when introducing the splitter, together with mode filters at each connection for keeping mode mixing under control (●), the key rate significantly decreases to 15 kb/s at an optical budget (OB) of 0 dB. For the same OB of 10 dB, this corresponds to a penalty of 23 dB in key rate compared to the straight-line ODN. Note that OB denotes the end-to-end loss allowance between the transmitter (Alice) and receiver (Bob), including all passive components and fiber spans. By removing M2 (■), the QBER improves by about 3%, but this comes at the cost of a 5 kb/s reduction in key rate. This removal creates a more unstable scenario that makes it more difficult to manage mode power uncertainties, thereby reducing the key rate. Furthermore, without M2, the system tolerates 1 dB less of transmission loss. This might be perceived as counterintuitive, but there is a trade-off between "cleaner" mode (lower QBER) and "stronger" signal (higher count rate). The optimal operating point is reached when a modest amount of mode filtering suppresses the dominant error mechanism without imposing excessive loss. Practically, this optimum can be tuned via the bend radius/number of turns (i.e., the strength of M2) and verified by measuring SKR as a function of the induced insertion loss. By placing M1 (▲), the QBER enhances by 0.7% and the key rate is increased by approximately 10 kb/s. This suggests that it is more advantageous to place the mode filter at the end of the channel, rather than at the splitter junction, and to avoid filtering the modes again prior to detection. Although filtering at the splitter can be beneficial, the modes will evolve differently along the fiber, potentially decreasing the QBER and not effectively coupling power to the LP01 mode.

We further investigated the robustness of the shortwave-QKD system under co-existence with 50 classical data channels from 1307.2 nm to 1618.7 nm (Fig. 3) modulated with 10 Gb/s/λ OOK data, having an aggregated launch power of 10.1 dBm. Quantum and classical channels are multiplexed at Alice's and Bob's sites using co-existence filters (π). These components consist of (i) a cleaning filter to suppress spontaneous emission tails, implemented via a wideband 850/1310 waveband splitter at Alice's side, and (ii) an 850/1310 separator, along with a cascade of wideband 850/1310 demultiplexer and a free-space bandpass filter (BPF) with a 10 nm FWHM bandwidth centered at 845 nm, integrated into Bob's polarimeter to suppress Raman noise. Figure 3 proves that the spontaneous emission tail (acquired with only one 850/1310 waveband separator) appears only above a spectral detuning of -42 THz, where they are filtered by Bob's BPF. Figure 4b shows the raw-key rate and QBER as a function of the OB for shortwave-QKD propagation in co-existence with 50 classical channels over an ODN including a 1×4 splitter, a 1 km long fiber and mode filter M2, which proved to be the most promising layout according to the previous characterization. The presence of classical channels considerably affects the key rate, reducing it to 11 kb/s and resulting in a 7.5 dB lower loss tolerance than in the absence of classical signals (Fig. 4b). The impact on the QBER highlights the sensitivity of QKD to few-mode conditions, though overall trends remain consistent regardless of the presence of classical channels.

In our experiment the launch of the QKD signal has been carefully optimized for all ODN configuration that furnish a PLC splitter. Few-mode speckle patterns at splitter junctions cause significant variation in transmission loss across its drop ports, as shown in Fig. 5a. This speckle-selective loss is highly sensitive to small wavelength detuning, which alter the speckle pattern, thus impacting the QKD performance. For example, Fig. 4b shows that QKD over drop port 1 has a 9 dB higher loss tolerance than port 4, which surpasses the QBER threshold at only 4 dB of OB. Similarly, port 1 achieves a raw-key rate of 14 kb/s, while port 4 is limited to 6 kb/s. These discrepancies are due to imbalanced power distribution and differential delay between modes [5,6] that lead to depolarization for our QKD setup, with even minor wavelength shifts for $\lambda_Q$ causing significant changes in the speckle pattern (Fig. 5b). Qualitatively, the dominant driver of the wavelength discrepancy of the speckle itself is the channel's modal/multipath interference, while the PLC accentuates port-to-port differences. Figure 5c shows the corresponding impact on the QBER, where a minor spectral detuning of just 3.5 pm can lead to a fluctuation of up to 5.86%, while under the worst spectral setup the key rate is negatively impacted up to 22 kb/s.

## 4. Shortwave-QKD over Field-Installed Fiber

To investigate a more realistic evolution of few-mode propagation, the 1-km long lab fiber was replaced by field-installed SMF28 fibers that were looped back to Bob at the lab (Fig. 2b). Two fiber loops were tested: one connecting the lab to an office over six fiber segments (KOS) spanning over 270 m, and a longer loop passing over the roof-top and having a total reach of 692 m over 12 segments (ROOF). The latter also includes a ~60-m long out-door cable duct, which has been exposed to weather effects (temperature of 12.8°C, wind speed of up to 44.3 km/h). In polarization-encoded operation, environmental SOP fluctuations can contribute to basis misalignment and QBER excursions. In addition, the exposure of the fiber link to environmental perturbations leads to variations of the collected spatial modes. The intermodal coupling driven by these perturbations alter the modal content and coupling efficiency, leading to a time-varying mode-coupling pattern unless adaptive mode launching techniques are employed. This impacts the received intensity and QBER, contributing to the performance differences in Fig. 5d. Instead, a comparison is provided over the in-door KOS loop. This link demonstrates high stability, achieving a peak-to-peak QBER excursion of 5.7% with a maximum QBER of 8.8% and an average raw-key rate of 19.2 kb/s, indicating the feasibility of long-term shortwave-QKD operation over installed fiber environments.

## 5. Conclusions

We have investigated the few-mode propagation of shortwave-QKD over (installed) standard telecom infrastructure. Especially PLC-based splitter components can lead to a large variation in key rate / QBER due to speckle-selective loss and differential mode delay along the drop span, which causes depolarization. The QBER can be improved by ~3% through mode filtering at the last optical joint and by optimizing the QKD mode launch conditions through spectral tuning. A secure-key rate of 12 kb/s has been accomplished in co-existence with 50 classical data channels.

**Funding.** This work has received funding from the EU Horizon Europe Program under projects QSNP (no. 101114043) and Qu-Test (no. 101113901).

**Data Availability:** Data underlying the results presented in this paper are not publicly available at this time but may be obtained from the authors upon reasonable request.